# Superconductivity of the hybrid Ruddlesden–Popper La$_5$Ni$_3$O$_{11}$ single crystals under high pressure

Mengzhu Shi[1,2,†], Di Peng[6,†], Kaibao Fan[1,2], Zhenfang Xing[7], Shaohua Yang[8], Yuzhu Wang[3], Houpu Li[1,2], Rongqi Wu[1,2], Mei Du[1,2], Binghui Ge[8], Zhidan Zeng[7], Qiaoshi Zeng[6,7,*], Jianjun Ying[1,5], Tao Wu[1,2,4,5,*], and Xianhui Chen[1,2,4,5,*]

1. CAS Key Laboratory of Strongly coupled Quantum Matter Physics, Department of Physics, University of Science and Technology of China, Hefei, Anhui 230026, China

2. Hefei National Laboratory for Physical Sciences at the Microscale, University of Science and Technology of China, Hefei, Anhui 230026, China

3. Shanghai Synchrotron Radiation Facility, Shanghai Advanced Research Institute, Chinese Academy of Sciences, Shanghai, China

4. Collaborative Innovation Center of Advanced Microstructures, Nanjing University, Nanjing 210093, China

5. Hefei National Laboratory, University of Science and Technology of China, Hefei 230088, China

6. Shanghai Key Laboratory of Material Frontiers Research in Extreme Environments (MFree), Institute for Shanghai Advanced Research in Physical Sciences (SHARPS), Shanghai, China

7. Center for High Pressure Science and Technology Advanced Research, Shanghai, China

8. Information Materials and Intelligent Sensing Laboratory of Anhui Province, Anhui Key Laboratory of Magnetic Functional Materials and Devices, Institutes of Physical Science and Information Technology, Anhui University, Hefei, China

†These authors contributed equally to this work
*Correspondence to: zengqs@hpstar.ac.cn, wutao@ustc.edu.cn, chenxh@ustc.edu.cn

**The discovery of high-temperature superconductivity in La$_3$Ni$_2$O$_7$ and La$_4$Ni$_3$O$_{10}$ under high pressure indicates that the Ruddlesden–Popper (*RP*) phase nickelates $R_{n+1}$Ni$_n$O$_{3n+1}$ (*R* = rare earth) is a new material family for high-temperature superconductivity. Exploring the superconductivity of other *RP* or hybrid *RP* phase nickelates under high pressure has become an urgent and interesting issue. Here, we report a novel hybrid *RP* nickelate superconductor of La$_5$Ni$_3$O$_{11}$. The hybrid *RP* nickelate La$_5$Ni$_3$O$_{11}$ is formed by alternative stacking of La$_3$Ni$_2$O$_7$ with**

n=2 and $La_2NiO_4$ with n=1 along the *c* axis. The transport and magnetic torque measurements indicate a density-wave transition at approximately 170 K near ambient pressure, which is highly similar to both $La_3Ni_2O_7$ and $La_4Ni_3O_{10}$. With increasing pressure, high-pressure transport measurements reveal that the density-wave transition temperature ($T_{DW}$) continuously increases to approximately 210 K with increasing pressure up to 12 GPa before the appearance of pressure-induced superconductivity, and the density-wave transition abruptly fades out in a first-order manner at approximately 12 GPa. The optimal superconductivity with $T_c^{onset}$ = 64 K and $T_c^{zero}$ = 54 K is achieved at approximately 21 GPa. On the other hand, high-pressure X-ray diffraction experiments reveal a structural phase transition from an orthorhombic structure to a tetragonal structure at approximately 4.5 GPa. In contrast to $La_3Ni_2O_7$ and $La_4Ni_3O_{10}$, the pressure-induced structural transition has no significant effect on either the density-wave transition or the superconductivity, suggesting a minor role of lattice degree of freedom in $La_5Ni_3O_{11}$. The present discovery extends the superconducting member in the *RP* nickelate family and sheds new light on the superconducting mechanism.

Since the discovery of superconductivity in cuprates, exploring high-temperature superconducting materials with similar crystal and electronic structures has become an important research direction[1,2,3,4,5,6,7]. A major breakthrough in this field was made in the infinite-layer nickelate $Nd_{0.8}Sr_{0.2}NiO_2$ thin films with a superconducting transition temperature ($T_c$) of 9--15 K in 2019[8]. Motivated by this groundbreaking finding, the Ruddlesden–Popper (*RP*) phase nickelates $R_{n+1}Ni_nO_{3n+1}$ (*R* = rare earth) with n = 2 and n = 3 are reported to exhibit superconductivity under pressure[9,10,11,12,13], which largely expands the family of nickelate superconductors. In these *RP* nickelates $R_{n+1}Ni_nO_{3n+1}$ (*R* = rare earth), the multilayer perovskite structure $(RNiO_3)_n$ is believed to be the fundamental building block for superconductivity. Clarifying the role of the multilayer perovskite structure $(RNiO_3)_n$ in the superconducting phase is important for building a theoretical model for the superconducting mechanism. At ambient pressure, the $NiO_6$ octahedron in the $(RNiO_3)_n$ structure is distorted and tilted, which leads to an orthorhombic structure. Moreover, in such an orthorhombic structure phase, a density-wave (DW) transition is widely observed at approximately 130--150 K in *RP* nickelates with n = 2 and 3, which involves both spin and charge density wave orders[14,15,16]. With increasing pressure, the distortion and tilting of the $NiO_6$ octahedron are strongly suppressed, and a pressure-induced structural transition from an orthorhombic structure to a tetragonal structure occurs at approximately

15 GPa[9,12,11]. Previous high-pressure transport measurements suggest that the DW transition is also suppressed with increasing pressure, and the pressure-dependent phase diagram of superconductivity and density-wave order suggests a possible competing scenario with a second-order manner [9,17].

In addition to RP-phase nickelates, hybrid *RP*-phase nickelates has been also reported[18,19,20]. Hybrid *RP*-phase nickelates are formed by alternative stacking of different *RP* phases along the *c* axis. To date, two hybrid *RP*-phase nickelates have been reported: the 1313 phase, with a chemical formula of $La_3Ni_2O_7$, and the 1212 phase, with a chemical formula of $La_5Ni_3O_{11}$[18,19,20]. In the 1313 phase, the $(LaNiO_3)_3$ layer and $La_2NiO_4$ layer alternately stack along the *c* axis and are separated by the LaO layer. Previous high-pressure transport measurements suggest possible high-temperature superconductivity in the 1313 phase, with an onset transition temperature of approximately 80 K[20]. Since there are no reports on superconductivity in the $La_2NiO_4$ phase, the superconducting pairing should come from the $(LaNiO_3)_3$ layer. This result further supports the use of the multilayer perovskite structure $(RNiO_3)_n$ as the fundamental building block for superconductivity. However, there is a hot debate on the superconducting phase for pressurized superconductors with the chemical formula $La_3Ni_2O_7$ and the onset transition temperature $T_c^{onset}$ = 80 K because this chemical formula can share either the *RP* phase $R_{n+1}Ni_nO_{3n+1}$ with n = 2 or the hybrid *RP* 1313 phase, which is formed by alternative stacking of the $(LaNiO_3)_3$ layer and the $La_2NiO_4$ layer along the *c* axis. Furthermore, $T_c^{onset}$ = 80 K in the hybrid *RP* 1313 phase seems to conflict with the reported superconductivity with a $T_c^{onset}$ of less than 30 K in the pressurized *RP* trilayer nickelate $La_4Ni_3O_{10}$[10,11,12,13]; more experiments on the origin of superconductivity in the 1313 phase are needed. Because the multilayer perovskite structure $(RNiO_3)_n$ serves as the fundamental building block for superconductivity, high-temperature superconductivity should be expected in the 1212 phase under pressure. In the 1212 phase, as shown in Fig. 1a, the single-layer and bilayer blocks of $NiO_6$ octahedron alternately stack along the *c* axis, forming the so-called hybrid *RP* 1212 nickelate[18]. In this work, we perform a systematic study on the pressure-dependent evolution of the electronic state in a hybrid *RP* 1212 nickelate single crystal with a chemical formula of $La_5Ni_3O_{11}$. High-pressure transport measurements using helium gas as the pressure transmitting medium revealed an unambiguous superconducting transition above ~ 12 GPa. The optimal superconducting transition temperature of a $T_c^{onset}$ of ~ 64 K and a zero-resistivity temperature ($T_c^{zero}$) of ~ 54 K is achieved at approximately 21 GPa. In addition, the pressure-dependent evolution of the superconductivity, DW transition and structure are also mapped out.

**Structure and density-wave transition at nearly ambient pressure**

The hybrid *RP* 1212 nickelate single crystal was synthesized through a melt–salt method (see Methods for details). After the flux was dissolved in water, the product was filtered with 400-mesh (~38.5 μm) sieves. A single crystal with typical dimensions of 0.1×0.1×0.02 mm was carefully checked via a four-circle diffractometer. Fig. 1a shows the crystal structure model of the as-grown hybrid *RP* 1212 nickelate single crystal (left panel) determined from single-crystal X-ray diffraction data, where single-layer and bilayer perovskite-like $NiO_6$ octahedrons alternately stack along the *c*-axis direction, as previously reported[18]. The space group is determined to be *Cmmm*, which is different from the previously reported *Immm*[18]. We note that the similar compound "1313" phase nickelate, where single-layer and trilayer blocks of $NiO_6$ octahedrons stack alternately along the *c*-axis direction, also adopts a *Cmmm* space group[19]. The structure of the hybrid *RP* 1212 nickelate is also confirmed by atomically resolved scanning transmission electron microscopy (STEM) images, where the alternate stacking of single-layer and bilayer blocks of $NiO_6$ octahedrons is clearly observed in Fig. 1b. The overlaid crystal structure model fits well with the STEM-HAADF image (Fig. 1b left panel). By grinding several pieces of hybrid *RP* 1212 nickelate single crystals, powder X-ray diffraction (XRD) patterns were collected at the Shanghai Synchrotron Radiation Facility at a wavelength of 0.4834 Å at moderate pressure (1.2 GPa) using helium gas as the pressure transmitting medium. With the Rietveld refinement method, the powder XRD pattern can be well fitted with the structural model solved from the single-crystal XRD data. No other *RP* phase was observed in the powder XRD pattern. In the hybrid *RP*-phase nickelate $La_5Ni_3O_{11}$, the out-of-plane Ni-O-Ni angle between the $NiO_6$ octahedrons is symmetry constrained to 180° (see Fig. 1a and Table S1), which is different from the value of 168° in $La_3Ni_2O_7$ with the *Amam* space group at ambient pressure. The out-of-plane Ni-O-Ni angle was previously thought to be critical for interlayer coupling between NiO planes, which favours superconductivity under high pressure. More detailed crystal data, structure refinements and bond angles are shown in Extended Data Table S1 and Extended Data Fig. 1. To obtain good electric contact, the temperature-dependent resistivity curve (*R*(*T*)) for the as-grown microcrystal was measured on a DAC with a small pressure (~3.5 GPa) (see the inset of Fig. 1d). As shown in Fig. 1d, the resistance curve (*R*(*T*)) exhibits a large hump at approximately 170 K, which is consistent with a previous report on powder samples at ambient pressure[18]. The anomaly in the resistivity curve is possibly due to a DW transition, which is similar to the other *RP* phase nickelates and the hybrid *RP* "1313" phase nickelate[14,15,19]. Magnetic torque measurements conducted on the hybrid *RP* 1212 nickelate microcrystal (Fig. 1e) confirmed a DW transition at ~170 K, which corresponds to the temperature at the maximum of the hump in the *R*(*T*) curve. We note that no obvious nonstoichiometry is observed on the basis of the EDX analysis (Extended Data Fig. 2) and the refinement of the single-crystal XRD data (see Methods).

**Pressure-induced superconductivity**

The electrical transport properties of the hybrid *RP* 1212 nickelate under various pressures were collected on a DAC using helium gas as the pressure transmitting medium[11]. Notably, the homogeneity of the pressure environment is very important for electrical transport measurements under pressure, especially for the hybrid *RP* 1212 nickelate. Owing to the large volume shrinkage of helium gas under high pressure, good electric contact is quite challenging in practice, and realistic electric contacts usually work well only within a limited pressure range. Here, we successfully measure the electrical transport in different pressure ranges on three pieces of $La_5Ni_3O_{11}$ single crystals, S2, S3 and S4, which are selected from the same batch. The electric contacts for sample S2 are good only for electrical measurements at relatively low pressures (below ~ 15 GPa), and the electric contacts for samples S3 and S4 are good only for electrical measurements at relatively high pressures (above ~ 15 GPa). For sample S2, the resistance at room temperature gradually decreases at a relatively low pressure (8.9–10.8 GPa, Fig. 2a), and the overall temperature-dependent behavior is similar to that at 3.5 GPa (Fig. 1d). When the applied pressure further increases above 11.7 GPa, although the DW transition temperature remains almost unchanged, the signature of the DW transition in transport is strongly suppressed and completely fades above ~ 13 GPa (Fig. 2b). This result suggests a pressure-induced first-order phase transition for the DW order (Fig. 4a). At ~ 11.7 GPa, a sharp drop in the *R*(*T*) curve indicates the emergence of superconductivity, with a $T_c^{onest}$ of 17.9 K (Fig. 2b). Above 11.7 GPa, the value of $T_c^{onest}$ continuously increases and reaches the optimal superconductivity, with the highest $T_c^{onest}$ value of ~ 64 K occurring at ~ 21.5 GPa (Figs. 2c and 2d, sample S3). As shown in Fig. 2d, there is a step-like transition at approximately 50 K due to possible inhomogeneity of the pressure environment, which leads to a zero-resistance temperature of only 42 K in sample S3 (Extended Data

Fig. 4a). By improving the homogeneity of the pressure environment, we finally obtain a $T_c^{onest}$ of ~ 64 K and a $T_c^{zero}$ of ~ 54 K for sample S4 (Fig. 2e), which are among the highest reported zero-resistance temperatures and the sharpest superconducting transitions for nickelate superconductors. As shown in Fig. 2e and Extended Data Fig. 4, we studied the superconducting transition under different magnetic fields perpendicular to the ab planes. The upper critical field ($H_{c2}$) is extracted with different criteria. As shown in Fig. 2f, there is a positive curvature in the $H_{c2}$-$T_c$ curve, which cannot be explained by a single-band Ginzburg–Landau (GL) model. Here, we use a two-band model at the clean limit to fit the upper critical field, which works quite well and yields $H_{c2}$ values of 20--28 T at the zero-temperature limit[21]. In $La_3Ni_2O_7$ with the *Amam* space group, the upper critical field along the out-of-plane direction is ~180 T at 18.9 GPa [9], which is much greater than that in our case. This low upper critical field in the hybrid *RP* 1212 nickelate is also confirmed in another single-crystal sample (Extended data Fig. 4b). Above 20 GPa, the superconducting transition temperature starts to slightly decrease with increasing pressure up to 25.2 GPa.

**Structural transition under pressure**

To further understand the electrical transport behavior under pressure, we measured the powder XRD patterns under various pressures up to 30.5 GPa for the hybrid 1212 nickelate by grinding several pieces of microcrystals with helium gas as the pressure transmitting medium at the Shanghai Synchrotron Radiation Facility at a wavelength of 0.4834 Å. Fig. 3 and Extended Data Fig. 5 summarize the main results. At ambient pressure, the 1212 nickelate microcrystal adopts an orthorhombic structure with a space group of *Cmmm*, which is characterized by the splitting of the (020) and (200) diffraction peaks. With increasing pressure, the diffraction peaks of (020) and (200) gradually merged, which indicates a structural transition from the orthorhombic phase to the tetragonal phase below 5.8 GPa (Fig. 3b). In the crystals of $La_3Ni_2O_7$ with the *Amam* space group and $La_4Ni_3O_{10}$, the pressure at which the structure transitions into the tetragonal phase is approximately 14 GPa, which is much greater than that of the hybrid *RP* 1212 nickelate [9,12]. The refinement of the powder XRD pattern at 5.8 GPa gives a tetragonal phase structure with a space group of *P4/mmm* (Extended data Fig. 5b), which is similar to the case of the hybrid *RP* 1313 phase under high pressure[20]. The tetragonal phase structure is maintained at 30.5 GPa. More detailed evolution of the lattice parameters and cell volume are refined and shown in Figs. 3c-d, where the lattice parameters show a progressive decrease under pressure. A careful analysis of the evolution of the lattice parameters of the *a*- and *b*-axes indicates that the critical pressure for the structural transition is approximately 4.5 GPa (Fig. 3c).

**Pressure-dependent phase diagram**

In Fig. 4a, we summarize the results of high-pressure transport and XRD diffraction into a pressure-dependent phase diagram. As the pressure increases, the crystalline structure of the hybrid *RP* 1212 nickelate transitions from a low-pressure orthorhombic phase (*Cmmm*) to a high-pressure tetragonal phase (*P4/mmm*) at a critical pressure of ~ 4.5 GPa, which is much lower than that of $La_3Ni_2O_7$ and $La_4Ni_3O_{10}$ (~ 14 GPa)[9,11]. In contrast to previous high-pressure transport measurements on $La_3Ni_2O_7$ and $La_4Ni_3O_{10}$, the density-wave transition in 1212 nickelate is quite robust during the structural transition and is continuously enhanced with increasing pressure. Notably, previous muon spin rotation (μSR) and nuclear magnetic resonance (NMR) experiments on pressurized $La_3Ni_2O_7$ revealed a pressure-enhanced spin-density-wave (SDW) transition[14,15]. We speculate that the DW transition in the hybrid *RP* 1212 nickelate is also related to a similar SDW transition, which needs further experimental investigation in the future. Above 11.7 GPa, the superconducting phase emerges with dome-like pressure-dependent behavior. Our present results indicate strong competition between possible SDW order and superconductivity. They are connected via a first-order phase transition in the pressure-dependent phase diagram. Finally, we also studied the relationship between $T_c$ and the average in-plane lattice ($a_p = \frac{1}{2}\sqrt{a^2 + b^2}$) in the hybrid *RP* 1212 nickelate. As shown in Fig. 4b, the relationship between the $T_c$ and the average in-plane lattice parameter in 1212 nickelate is similar to that in pressurized $La_3Ni_2O_7$ [9,22], in which pressure-induced superconductivity appears in the structure with a relatively small $a_p$ (< 3.77 Å). This result suggests that the multilayer perovskite structure ($RNiO_3$)$_3$ is the fundamental building block for superconductivity. Very recently, by utilizing compressed strain through a substrate, ambient-pressure superconductivity has been observed in $La_{3-x}Pr_xNi_2O_7$ films [23,24]. The relationship between $T_c$ and the average in-plane lattice in these $La_{3-x}Pr_xNi_2O_7$ films also shows a similar behavior as that of the bulk samples under pressure. Here, the observation of pressure-induced superconductivity in the hybrid *RP* 1212 nickelate suggests an alternative route to achieve ambient-pressure superconductivity in the hybrid *RP* nickelates. The average in-plane lattice parameter of $La_2NiO_4$ is approximately 3.85–3.87 Å, which is relatively larger than the average in-plane lattice parameter of $La_3Ni_2O_7$ ($a_p$ =3.835 Å) [25]. If we can replace the $La_2NiO_4$ layer with another *RP* layer with a smaller $a_p$, it might be possible to tune the value of $a_p$ to the superconducting region, as shown in Fig. 4b. This deserves further experimental exploration of new hybrid *RP* nickelates.

In summary, by performing high-pressure transport measurements with helium gas as the pressure transmitting medium, we discovered pressure-induced superconductivity with an optimal $T_c^{onest}$ of ~ 64 K in a hybrid RP 1212 nickelate single crystal with the chemical formula $La_5Ni_3O_{11}$. In contrast to previously reported pressure-induced superconductivity in $La_3Ni_2O_7$ and $La_4Ni_3O_{10}$, the ambient-pressure DW order in the hybrid RP 1212 nickelate is quite robust against pressure, and the pressure-dependent phase diagram demonstrates a first-order phase transition between the low-pressure DW order and the high-pressure superconductivity. In addition, a structural transition from the orthorhombic phase to the tetragonal phase is also revealed at 4.5 GPa. Finally, this work also suggests the potential for realizing ambient-pressure superconductivity via sophisticated structural design of hybrid *RP* nickelates. The present discovery extends the superconducting member in the *RP* nickelate family and sheds new light on the superconducting mechanism.

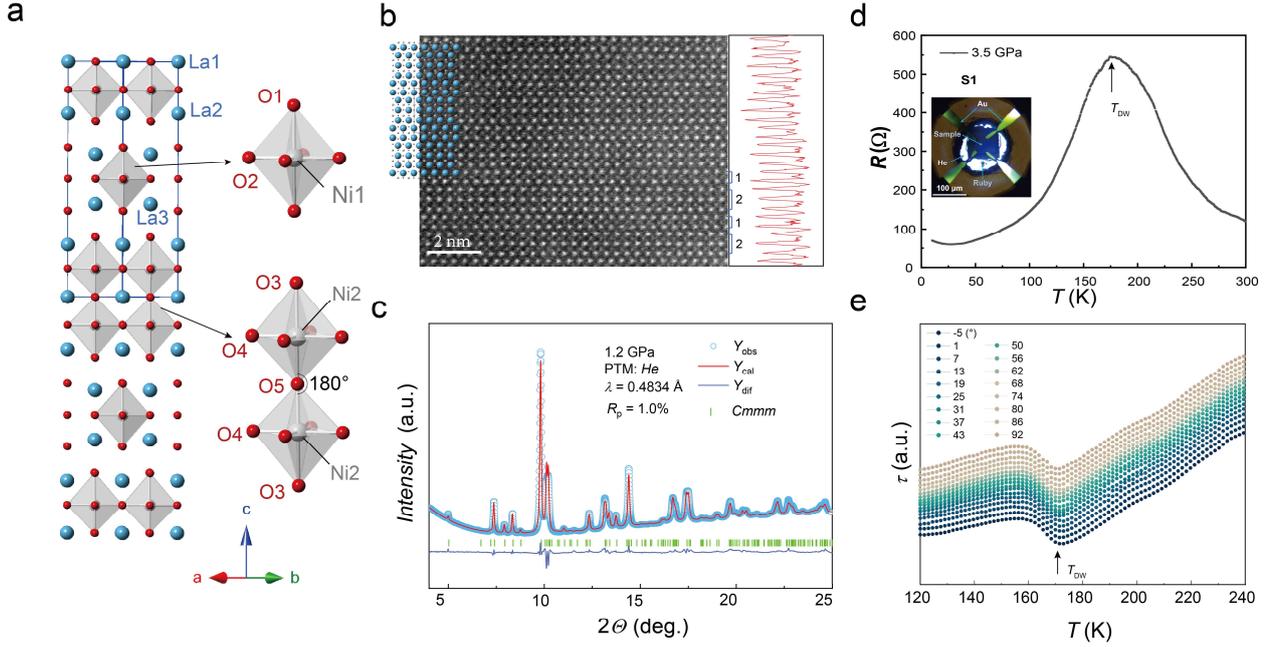

**Figure 1. Structure and physical properties of $La_5Ni_3O_{11}$.** **(a):** Crystal structure model of $La_5Ni_3O_{11}$ (left panel) and the stacking units of the monolayer and bilayer $NiO_6$ octahedrons (right panel) solved from the single-crystal X-ray diffraction data. The bond angle of Ni2-O-Ni2 in the bilayer subslab along the $c$-axis direction is 180°. **(b):** Cross-sectional STEM image of $La_5Ni_3O_{11}$ along the [110] direction. There are clear monolayer (denoted as '1') and bilayer (denoted as '2') subslab stacking along the $c$-axis direction. The overlaid crystal structure model fits well with the STEM-HAADF image (left panel). The right panel shows the line intensity profile for the image shown in the left panel. **(c):** Powder XRD pattern (blue circles) collected by gridding several microcrystals of $La_5Ni_3O_{11}$ at a moderate pressure of 1.2 GPa with a wavelength of 0.4834 Å. Adopting the Rietveld refinement method, the powder XRD pattern can be well fitted (red lines) via the structural model shown in (a). The blue lines indicate the difference between the observed and calculated data. The short green vertical lines indicate the calculated diffraction peak positions. **(d):** The temperature-dependent resistivity curve of $La_5Ni_3O_{11}$ at a small pressure (3.5 GPa) on the DAC when helium gas is used as the pressure transmitting medium. The inset shows the sample connected with the gold electrodes inside the gasket hole. There is a large hump at approximately 170 K in the $R(T)$ curve, which resembles that of the previously reported electrical transport data collected on powder samples at ambient pressure. **(e):** Temperature-dependent magnetic torque data ($\tau(T)$) for $La_5Ni_3O_{11}$ at various angles. There is a kink at approximately 170 K in the $\tau(T)$ curve, which is consistent with the anomaly in the $R(T)$ curve shown in (d). These results indicate a possible DW transition in $La_5Ni_3O_{11}$ at approximately 170 K.

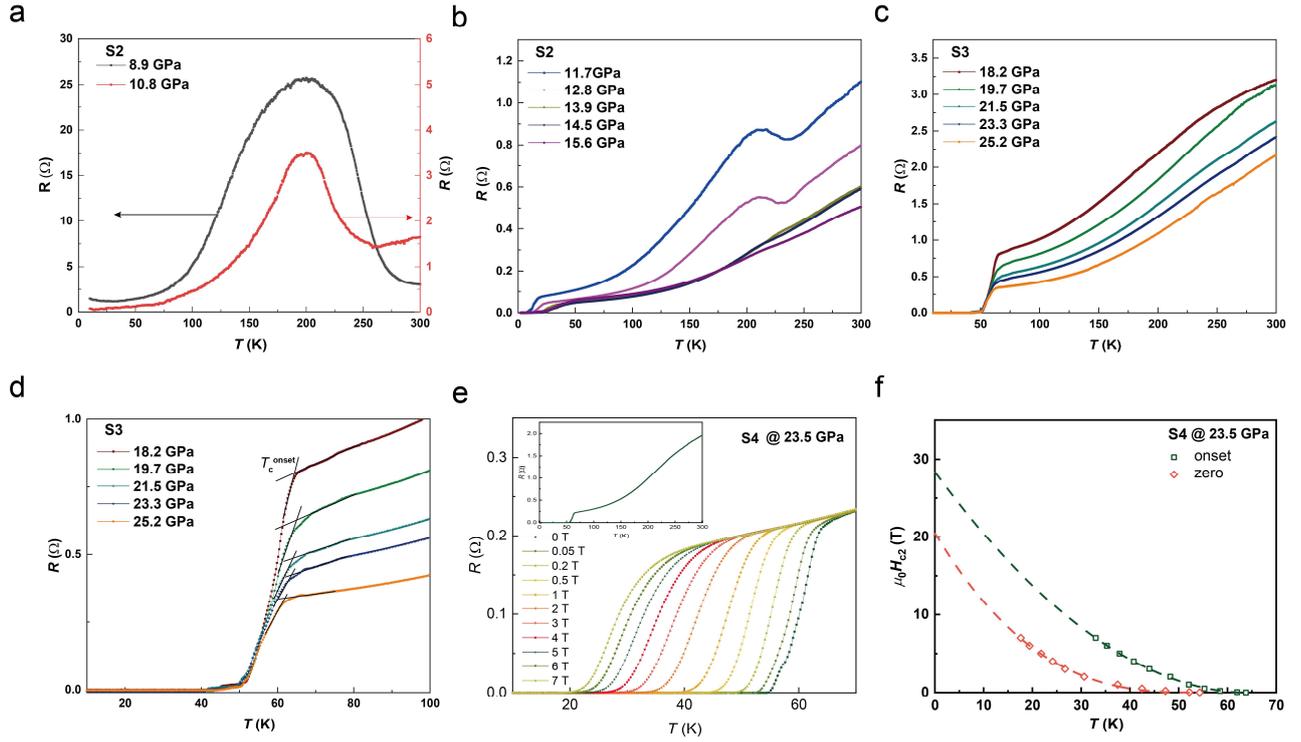

**Figure 2. Electrical transport properties of $La_5Ni_3O_{11}$ single crystals under various pressures for three samples, S2, S3 and S4. (a):** $R(T)$ curves for $La_5Ni_3O_{11}$ (S2) in a relatively lower pressure range (8.9–10.8 GPa). **(b):** $R(T)$ curves for $La_5Ni_3O_{11}$ (S2), where superconductivity begins to occur. **(c):** $R(T)$ curves of $La_5Ni_3O_{11}$ (S3) with zero resistance at ~ 40 K at pressures above 18.2 GPa. **(d):** Enlarged view of (c), where the $T_c^{onset}$ is defined graphically. **(e):** $R(T)$ curves for $La_5Ni_3O_{11}$ (S4) under various magnetic fields along the $c$-axis direction. The onset $T_c$ is quickly suppressed to a lower temperature with increasing magnetic field. The inset shows the $R(T)$ curve at 23.5 GPa without applying the magnetic field for the $La_5Ni_3O_{11}$ crystal (S4), which has similar electrical transport behavior to that of sample S3 and shows a $T_c^{onset}$ at ~64 K and a $T_c^{zero}$ at ~54 K. **(f):** The upper critical field extracted from (e). There is an obvious positive curvature in the $H_{c2}$-$T_c$ curve. The upper critical field at the zero-temperature limit is fitted via the two-band model at the clean limit with the equation $H_{c2}(T) = H_{c2}(0) \times (1 - (T/T_c))^{1+\alpha}$, where $H_{c2}(0)$ and $\alpha$ are fitting parameters[21].

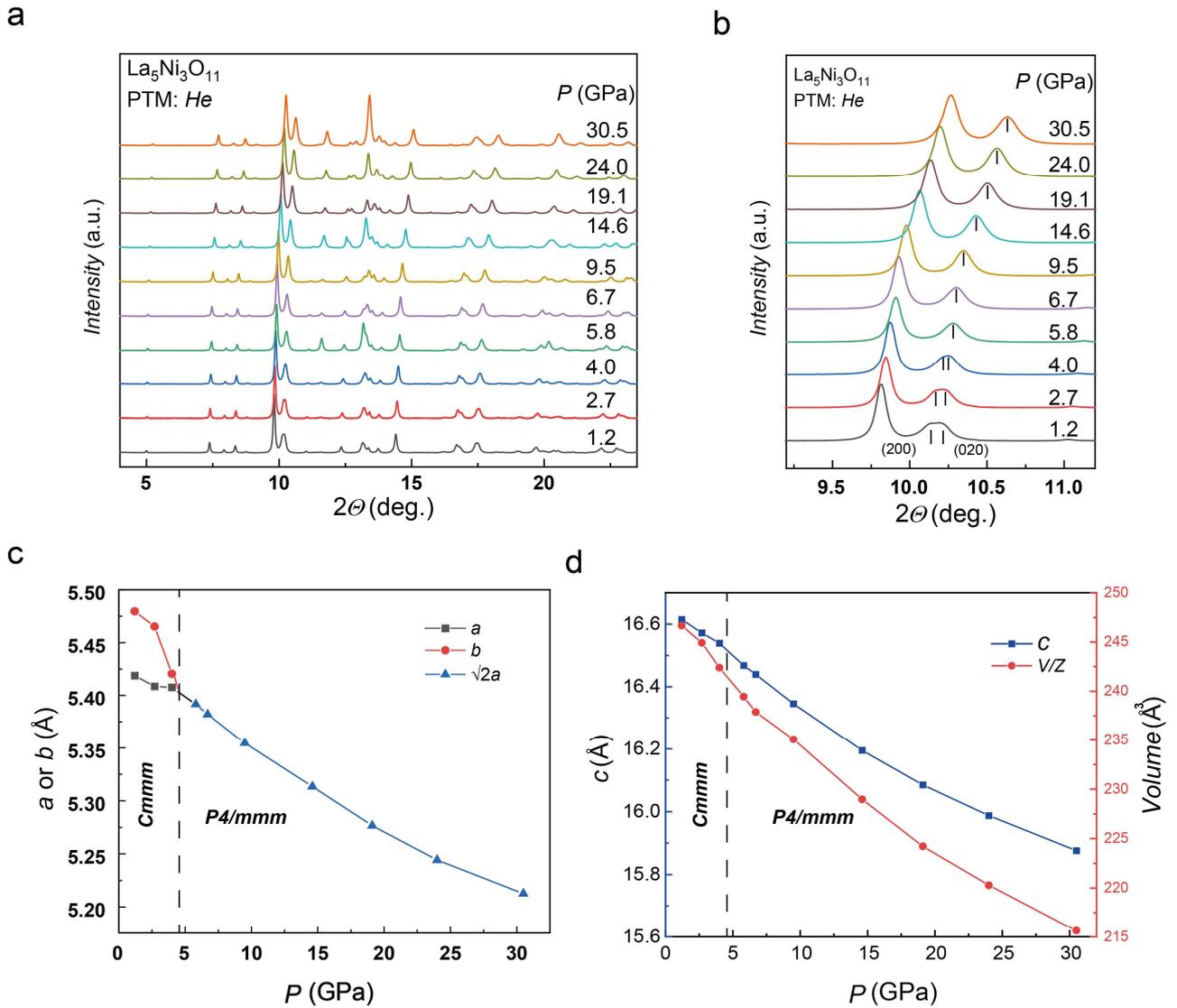

**Figure 3. Structural evolution of the $La_5Ni_3O_{11}$ single crystal with pressure. (a):** Powder XRD patterns of $La_5Ni_3O_{11}$ under various pressures. **(b):** Enlarged view of (a) at the $2\theta$ range of 9.2–11.2°, where the diffraction peaks of (200) and (020) gradually merged with increasing pressure. **(c):** Calculated lattice parameters of the *a*- and *b*-axes for $La_5Ni_3O_{11}$ under various pressures. The values of *a* and *b* decrease quickly and become equal at approximately 4.5 GPa (dashed line). **(d):** Calculated lattice parameters of the *c*-axis and specific cell volume for one $La_5Ni_3O_{11}$ molecule (*V/Z*) under various pressures, where *V* is the cell volume and *Z* is the molecule number for one crystal cell. The dashed line is located at 4.5 GPa.

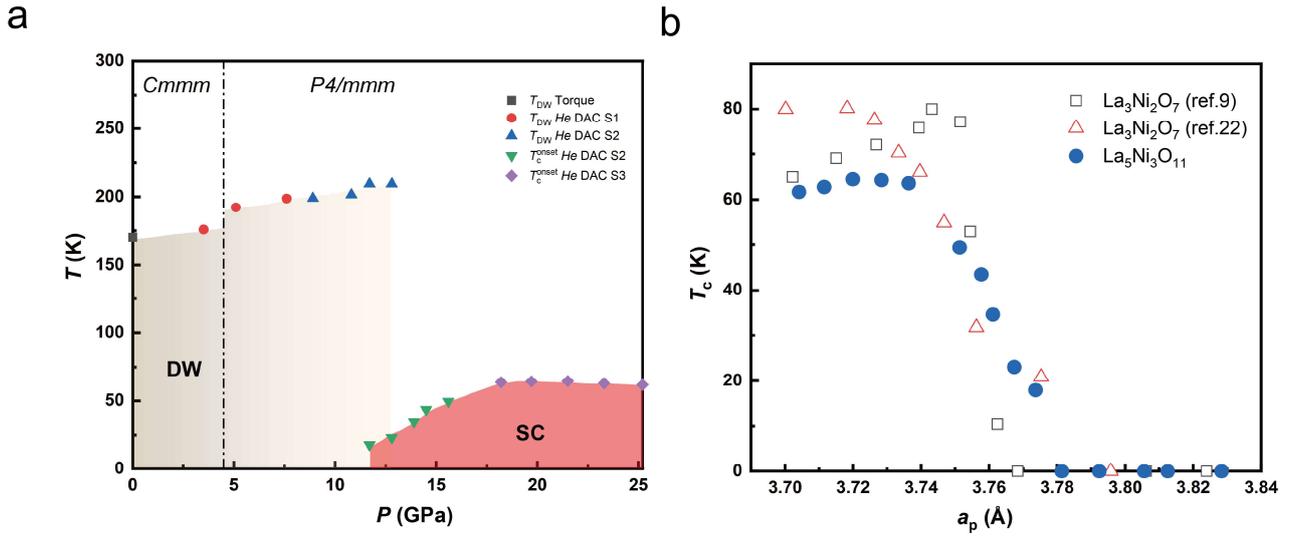

**Figure 4. Phase diagram of the hybrid *RP* 1212 nickelates with the chemical formula La$_5$Ni$_3$O$_{11}$. (a):** Pressure-dependent crystal structure, DW transition and $T_c^{onset}$ for the La$_5$Ni$_3$O$_{11}$ single crystal. The dotted dashed line separates the crystal structures of *Cmmm* and *P4/mmm*. The gray and red areas indicate the density wave regime (DW) and superconductivity regime (SC), respectively. **(b):** The $T_c^{onset}$ and the average in-plane lattice ($a_p$) relationship for La$_5$Ni$_3$O$_{11}$.

**Method**

**Sample growth:** The $La_5Ni_3O_{11}$ crystals were grown via a melt salt method. First, the $La_5Ni_3O_x$ precursor (P) was obtained via a standard sol–gel process. Specifically, the La source (lanthanum nitrate hexahydrate), Ni source (nickel (II) nitrate hexahydrate) and complexing agent (citric acid, CA) were dissolved in water at a molar ratio of La:Ni:CA= 5:3:8. The above solution was preheated at 140 °C for approximately 24 h to obtain a dry gel, which was then transferred into a muffle furnace where the temperature was slowly increased to 400 °C and maintained for another 10 h. Second, the above precursor (P) was mixed with a salt flux (NaCl/KCl mixture) at a mass ratio of P:NaCl:KCl = 1:14:16 and loaded into a corundum crucible. The corundum crucible was heated to 1150 °C for 10 h, maintained at this temperature for 48 h, and then slowly cooled to 1110 °C within 7 days. Microcrystals with a typical size of 0.1×0.1×0.02 mm were obtained after the flux was washed with water.

**Structural and composition characterization at ambient pressure:** The as-grown microcrystal was mounted on the sample holder using high-vacuum silicon grease as the glue. Single-crystal X-ray diffraction (SC-XRD) data were collected on a four-circle diffractometer (Rigaku, XtaLAB PRO 007HF) with Cu $K\alpha$ radiation at the Core Facility Center for Life Sciences, USTC. The structure was solved and refined via Olex-2 with the ShelXT and ShelXL packages. The detailed structural data are shown in Table S1. All the crystals were first checked via a four-circle diffractometer before they were used to conduct further physical measurements. Energy-dispersive X-ray spectroscopy (EDX) equipped with a scanning electron microscope (SEM, Hitachi SU8220) was used to characterize the chemical composition. The element ratio of La:Ni is approximately 1.67:1 (Extended Data Fig. 2). The refinement of the occupancy of the oxygen sites on the basis of the SC-XRD data gives a value of 0.984--1.072, which indicates nearly full occupation of all these oxygen sites. There is only one oxygen site (O3 site, see Fig. 1a) that is smaller than 1. These results indicate negligible nonstoichiometry in the as-grown $La_5Ni_3O_{11}$ single crystal. The scanning transmission electron microscopy (STEM) images were collected on a Thermo Fischer Scientific Titan Themis Z microscope with a working voltage of 300 kV.

**Magnetic torque measurement:** Using an SCL piezoresistive cantilever, torque magnetometry data were collected via a physical property measurement system (PPMS, Quantum Design Inc., DynaCool-14T). The sample was carefully attached to the tip of the cantilever, which was fixed on a horizontal rotator. The sample was rotated in the range of $\theta$ (the angle between the magnetic field vector $H$ (14T) and the flat plane of the $La_5Ni_3O_{11}$ crystal) from 0° to 90° under isothermal conditions.

**Electrical transport and XRD measurements under high pressure:** Resistance curves for the $La_5Ni_3O_{11}$ single crystals under high pressure were measured in a diamond anvil cell (DAC) using helium gas as the pressure transmitting medium. The pressure was applied, and the mixture was calibrated by shifting the ruby florescence at room temperature. The transport measurements were conducted in a Physical Properties Measurement System (PPMS-9, Quantum Design Inc.). The powder XRD data of $La_5Ni_3O_{11}$ under pressure were collected by gridding several pieces of microcrystals at the Shanghai Synchrotron Radiation Facility via an X-ray beam with a wavelength of 0.4834 Å. Helium gas was used as the pressure transmitting medium. The powder XRD data were refined via GSAS software to obtain the lattice parameters under different pressures.

## Data availability

The data that support the findings of this study are available from the corresponding author upon request.

## Code availability

The codes that support the findings of this study are available from the corresponding author upon request.

## Acknowledgements

We acknowledge fruitful discussions with Ho-kwang Mao, Zhengyu Wang and Ziji Xiang. We also thank Zhongliang Zhu, Fujun Lan, Yuxin Liu, and Hongbo Lou for their experimental assistance. This work is supported by the National Natural Science Foundation of China (Grant Nos. 12494592, 12488201, 11888101, 12034004, 12161160316, 12325403, and 12204448), the National Key R&D Program of the MOST of China (Grant No. 2022YFA1602601), the Chinese Academy of Sciences under contract No. JZHKYPT-2021-08, the CAS Project for Young Scientists in Basic Research (Grant No. YBR-048), and the Innovation Program for Quantum Science and Technology (Grant No.

2021ZD0302800). D.P. and Q.Z. acknowledge the financial support from the Shanghai Science and Technology Committee (Grant No. 22JC1410300) and Shanghai Key Laboratory of Material Frontiers Research in Extreme Environments (Grant No.22dz2260800). A portion of this research used resources at the beamline 17UM of the Shanghai synchrotron radiation facility.

## Author contributions

X.H.C. conceived the research project and coordinated the experiments. M.Z.S. grew the single crystals and performed the structural characterization at ambient pressure with the help of R.Q.W. and M.D.; H.P.L. and K.B.F. measured the magnetic torque data; S.H.Y. and B.H.G. collected the TEM images; D.P. performed the resistance measurements using helium gas as the pressure-transmitting medium under pressure with the help of Q.S.Z.; D.P., Z.F.X. and Y.Z.W. performed the synchrotron powder diffraction measurements and analysis under high pressure using helium gas as the pressure-transmitting medium with help from Q.S.Z. and Z.D.Z.; M.Z.S., D.P., J.J.Y., T.W. and X.H.C. analysed the data; M.Z.S., D.P., K.B.F., T.W. and X.H.C. wrote the paper with inputs from all the authors.

# Extended Figures and Tables

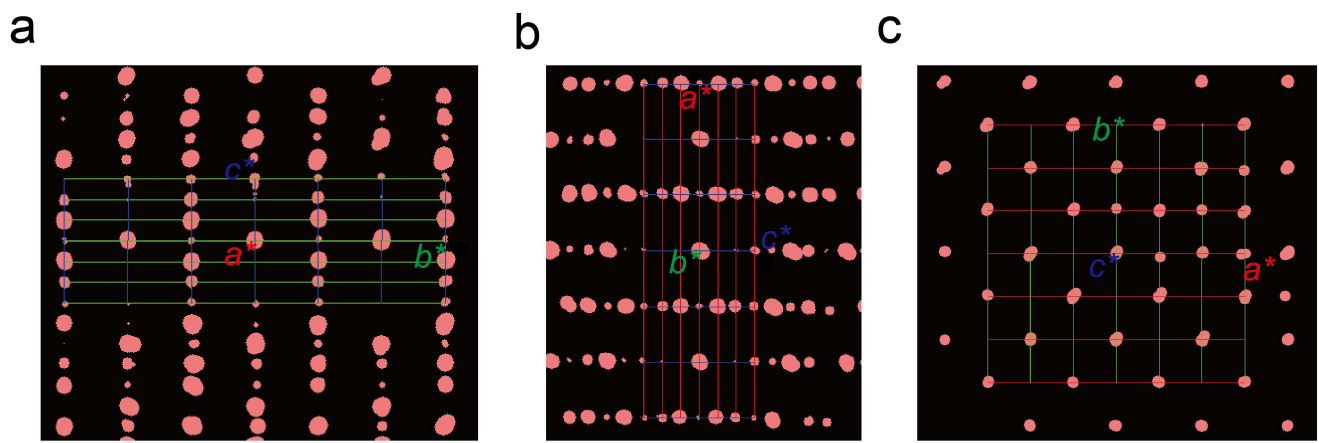

**Extended data Figure 1.** (a)-(c) Reciprocal lattice data of $La_5Ni_3O_{11}$ along the $a^*$, $b^*$ and $c^*$ axes, respectively. The size of the spots represents the intensity of the diffraction peaks.

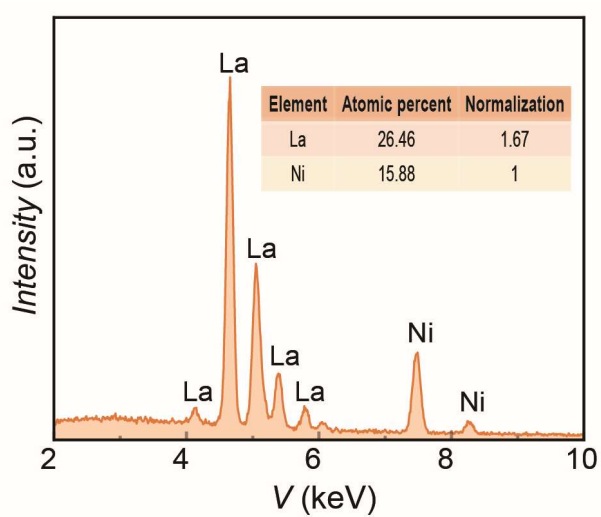

**Extended data Figure 2. EDX results of the as-grown microcrystals.** The element ratio is La:Ni = 1.67:1, which is consistent with the chemical formula of $La_5Ni_3O_{11}$.

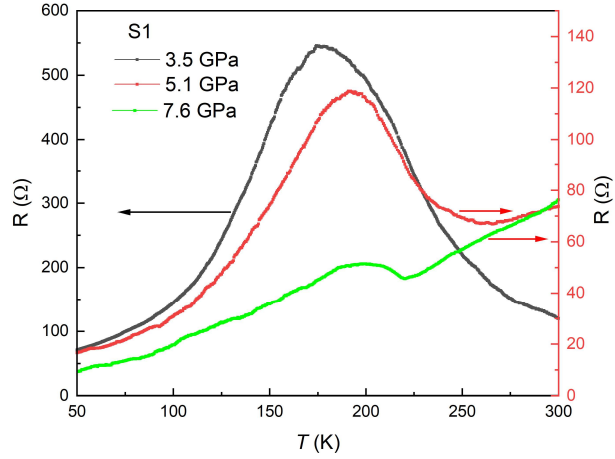

**Extended data Figure 3.** Temperature-dependent resistance curves for $La_5Ni_3O_{11}$ (S1) at various pressures.

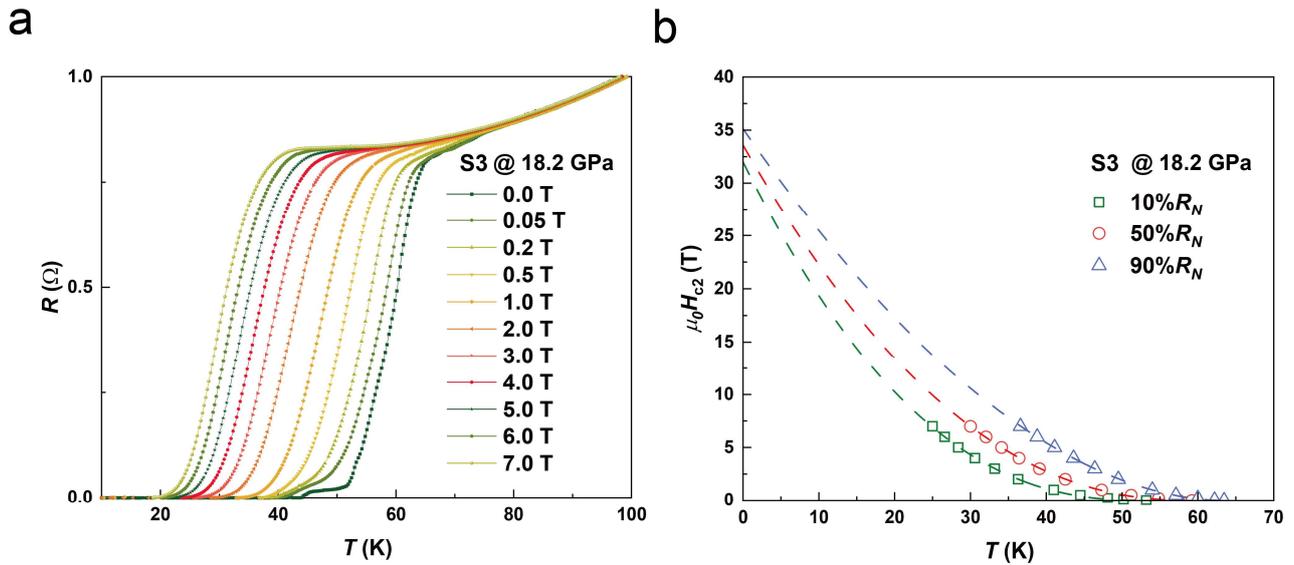

**Extended data Figure 4.** (a) $R(T)$ curves for $La_5Ni_3O_{11}$ (S3) at 18.2 GPa under various magnetic fields along the $c$-axis direction. The onset $T_c$ is quickly suppressed to a lower temperature with increasing magnetic field. (b) The upper critical field extracted with different criteria in (a), where the $R_N$ is the resistance at the normal state. There is an obvious positive curvature in the $H_{c2}$-$T_c$ curve. The upper critical field at the zero-temperature limit is fitted via the two-band model at the clean limit.

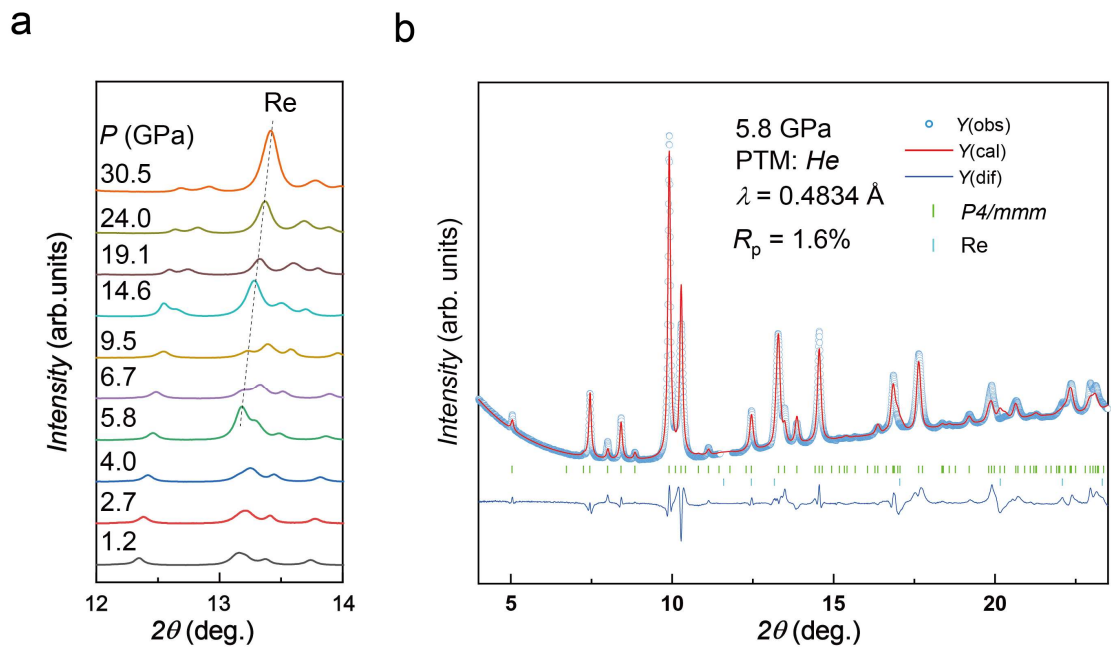

**Extended data Figure 5. Powder XRD patterns of the La$_5$Ni$_3$O$_{11}$ crystal under various pressures.** (a) Powder XRD patterns of the La$_5$Ni$_3$O$_{11}$ crystal under various pressures in the $2\theta$ range of 12–14°, where the dashed line indicates the signal from the gasket (Re). (b) Rietveld refinement of the powder XRD pattern for La$_5$Ni$_3$O$_{11}$ at 5.8 GPa. The collected data can be well fitted via the space group *P4/mmm*. The blue circles and red lines represent the observed and calculated data, respectively. The blue lines indicate the difference between the observed and calculated data. The short green and cyan vertical lines indicate the calculated diffraction peak positions of La$_5$Ni$_3$O$_{11}$ with *P4/mmm* space group and Re.

Table S1 Crystal data, structure refinement and bond angel for $La_5Ni_3O_{11}$.

| | |
|---|---|
| Identification code | LNO1212 |
| Empirical formula | $La_5Ni_3O_{11}$ |
| Formula weight | 1046.68 |
| Temperature/K | 300.4(4) |
| Crystal system | orthorhombic |
| Space group | *Cmmm* |
| *a*/Å | 5.4264(5) |
| *b*/Å | 5.4485(5) |
| *c*/Å | 16.5750(12) |
| $\alpha$/° | 90 |
| $\beta$/° | 90 |
| $\gamma$/° | 90 |
| Volume/Å$^3$ | 490.05(7) |
| Z | 2 |
| $\rho_{calc}$ g/cm$^3$ | 7.093 |
| $\mu$/mm$^{-1}$ | 171.053 |
| F(000) | 914 |
| Crystal size/mm$^3$ | 0.073 × 0.052 × 0.007 |
| Radiation | Cu K$\alpha$ ($\lambda$ = 1.54184) |
| 2$\Theta$ range for data collection/° | 10.674 to 156.012 |
| Index ranges | -6 ≤ h ≤ 6, -6 ≤ k ≤ 6, -21 ≤ l ≤ 20 |
| Reflections collected | 2823 |
| Independent reflections | 330 [$R_{int}$ = 0.0610, $R_{sigma}$ = 0.0303] |
| Data/restraints/parameters | 330/0/39 |
| Goodness-of-fit on $F^2$ | 1.112 |
| Final *R* indexes [I>=2$\sigma$ (I)] | $R_1$ = 0.0453, $wR_2$ = 0.1244 |
| Final *R* indexes [all data] | $R_1$ = 0.0463, $wR_2$ = 0.1253 |
| Largest diff. peak/hole / e Å$^{-3}$ | 3.97/-3.00 |
| In-plane Ni-O/Å | Ni1-O2: 1.922<br>Ni2-O4: 1.922 |
| Out-of-plane Ni-O/Å | Ni1-O1: 2.217<br>Ni2-O3: 2.236<br>Ni2-O5: 1.985 |
| In-plane Ni-O-Ni/deg | Ni1-O2-Ni1: 180<br>Ni2-O4-Ni2: 180 |
| Out-of-plane Ni-O-Ni/deg | Ni2-O5-Ni2: 180 |